\documentclass[10pt,twocolumn,prl]{revtex4-1}
\usepackage[T1]{fontenc}
\usepackage[latin9]{inputenc}
\usepackage{xcolor}
\usepackage[english]{babel}
\usepackage{amsmath}
\usepackage{amssymb}
\usepackage{graphicx}
\usepackage{ulem}
\usepackage[colorlinks]{hyperref}

\makeatletter
\global\long\def\ket#1{\left| #1\right\rangle }
\global\long\def\bra#1{\left\langle #1 \right|}

\global\long\def\abs#1{\left|#1\right|}

\global\long\def\t#1{\text{#1}}

\definecolor{applegreen}{rgb}{0.55, 0.71, 0.0}

\newcommand{\beginsupplement}{%
        \setcounter{table}{0}
        \renewcommand{\thetable}{S\arabic{table}}%
        \setcounter{figure}{0}
        \renewcommand{\thefigure}{S\arabic{figure}}%
        \setcounter{section}{0}
        \renewcommand{\thesection}{S\arabic{section}}%
        \setcounter{section}{0}
        \renewcommand{\thesection}{S\arabic{section}}%
        \setcounter{subsection}{0}
        \renewcommand{\thesubsection}{S\arabic{section}.\arabic{subsection}}%
        \setcounter{equation}{0}
        \renewcommand{\theequation}{S\arabic{equation}}%
     }
\makeatother

\begin{document}
\title{Fate of the Quasi-condensed State for Bias-driven Hard-Core Bosons in one Dimension}
\author{T. O. Puel$^{1,2}$, S. Chesi$^{3,4}$, S. Kirchner$^{5,6}$, P. Ribeiro$^{7,3}$}
\address{$^{1}$Zhejiang Institute of Modern Physics, Zhejiang University,
Hangzhou, Zhejiang 310027, China~\\
$^{2}$Department of Physics and Astronomy, University of Iowa, Iowa City, Iowa 52242, USA~\\
$^{3}$Beijing Computational Science Research Center, Beijing 100193,
China~\\
$^{4}$Department of Physics, Beijing Normal University, Beijing 100875,
China~\\
$^{5}$Department of Electrophysics, National Yang Ming Chiao Tung University, Hsinchu 30010, Taiwan\\
$^{6}$Center for Emergent Functional Matter Science, National Yang Ming Chiao Tung University, Hsinchu 30010, Taiwan\\
$^{7}$CeFEMA, Instituto Superior T\'{e}cnico, Universidade de Lisboa
Av. Rovisco Pais, 1049-001 Lisboa, Portugal
}
\begin{abstract}
Bosons in one dimension display a phenomenon called quasi-condensation, where correlations decay in a powerlaw fashion.
We study the fate of quasi-condensation in the non-equilibrium steady-state of a chain of hard-core bosons coupled to macroscopic leads which are held at different chemical potentials. 
It is found that a finite bias destroys the quasi-condensed state and the critical scaling function of the quasi-condensed fraction, near the zero bias transition, is determined. Associated critical exponents are determined and numerically verified.
Away from equilibrium, the system exhibits  exponentially decaying correlations that are characterized by a bias-dependent correlation length that diverges in equilibrium. 
In addition, power-law corrections are found, which are characterized by an exponent that depends on the chain-leads coupling and is non-analytic at zero bias.
This exactly-solvable nonequilibrium strongly-interacting system has the remarkable property that, the near-equilibrium state at infinitesimal bias, cannot be obtained within linear response.  
These results aid in unraveling the intricate properties spawned by strong interactions once liberated from equilibrium constraints. 
\end{abstract}
\maketitle

The properties of states of matter far from thermal equilibrium and their relation to their equilibrium counterparts has become a topical issue of quantum matter research. 
In equilibrium, it has long been recognized that thermodynamic phases of low dimensional systems
are generally suppressed, due to the enhanced role of quantum fluctuations. This understanding rests at least in part on the range of analytical methods that are available for one-dimensional (1D) systems.
For 1D hard-core bosons (HCB) this  amounts  to a phenomena commonly referred to as quasi-condensation where, instead of a macroscopic occupation of the condensate wave-function, the number of bosons in the ground-state increases as $\sqrt{N_b}$ where $N_b$ is the total number of bosons.  
Quasi-condensation is accompanied by off-diagonal quasi-long-range order, i.e., power-law decay of correlations, that characterises the quasi-condensed state. 
First discovered for homogeneous systems \cite{doi:10.1063/1.1704196,PhysRevLett.42.3}, quasi-condensed  states of 1D HCB  were shown to be ubiquitous, arising in the presence of  harmonic trapping  \cite{PhysRevA.70.031603}, periodic \cite{PhysRevA.63.033601,PhysRevA.67.041601,PhysRevA.67.043607} and even quasi-periodic potentials \cite{PhysRevA.87.043635}. 
The emergence of dynamic quasi-condensation has also been observed in some non-equilibrium closed systems
\cite{PhysRevLett.93.230404,PhysRevLett.115.175301,PhysRevLett.124.110603}. 
On the other hand, while quasi-condensation appears to be a persistent feature of 1D HCB,  
its fate in open systems far from equilibrium has so far not been addressed. 

This question is of significant interest, as the relation of closed vs open non-equilibrium steady states (NESS) is highly nontrivial. Systems with identical equilibrium behavior can deviate drastically as, e.g., the involved symmetries differ~\cite{Hohenberg.77,Mitra.06}.  Among non-equilibrium setups, typical transport configurations where the 1D systems is coupled to external reservoirs at different thermodynamic potentials are of great practical and theoretical relevance. For open quantum spin chains in the presence of a bias, it was recently shown that new behaviour can emerge, which is absent in equilibrium~\cite{PhysRevLett.122.197601,PhysRevB.103.035108}.
There, the ordered state is robust to small applied biases but transitions discontinuously to the disordered state at large bias, through a mixed order phase transition.  Interestingly, at that transition, the correlation length diverges. In contrast to the gapped ordered state of such spin-chains, the equilibrium quasi-condensed state is gapless, thus its response to any non-equilibrium drive might qualitatively differ~\cite{Ribeiro.15}.

These considerations motivate us to investigate a system of HCB in the presence of an applied bias and, in particular, address the fate of the quasi-condensed state.
Naturally, considerable attempts have been made to extend the analytical methods, available to one-dimensional equilibrium systems, to both closed and open systems far from equilibrium.  
For the closed case, these approaches include hydrodynamic methods for integrable models and generalized conformal field theory techniques \cite{Bernard_2016,Alba.21} for non-integrable ones. Approaches based on generalized Boltzmann-type equations \cite{Levchenko.11,Micklitz.11, Micklitz.12} were developed for dealing with open system dynamics.
Of particular relevance to the present work is the extension of the bosonization technique to nonequilibrium setups, pioneered in Refs.\  \cite{Gutman.08, Gutman.09,Gutman.11}.
Gutman et al.\  have shown that correlation functions, of interacting one-dimensional electrons, can be expressed through the asymptotics of Toeplitz determinants for imposed nonequilibrium electron distribution functions.
Similar considerations also apply to the bosonic Tonks-Girardeau gas \cite{Gutman.11}.

In this letter we obtain the steady-state properties of a chain of hard-core bosons coupled at its ends to leads in the wide-band limit.    
We show that the quasi-condensed state is unstable towards an applied bias and characterize the ensuing NESS in terms of its single-particle equal-time correlators.
We analyze the correlation length divergence in terms of the bias and determine how the quasi long-range order is restored. 
It is demonstrated that there are power-law corrections to the exponential decay, which depend on the bias in a non-analytic way once the thermodynamic limit is taken. As transport setups can now be readily engineered in confined ultracold atomic systems \cite{Brantut2012,Chien:2015tu,Krinner_2017}, a thorough understanding of quasi-condensation in open systems far from equilibrium is timely and topical. Such an understanding might also shed new light on the similarities, and differences, between non-equilibrium transport in cold atoms and condensed matter setups. 

\begin{figure}
\hfill{}\includegraphics[width=0.99\columnwidth]{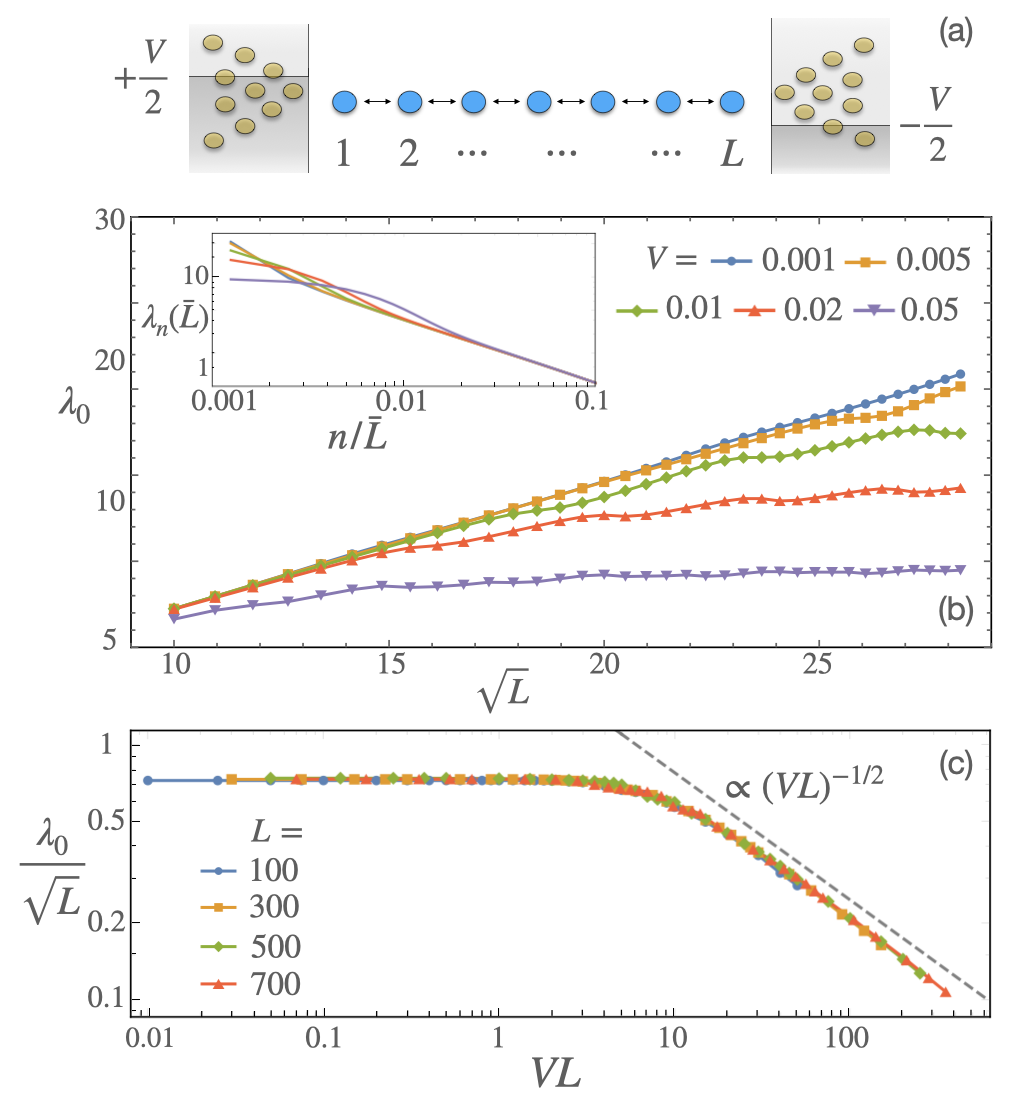}\hfill{}
\caption{
(a) Sketch of the HCB chain coupled to bosonic reservoirs. 
(b) Maximum natural-orbital occupation $\lambda_0$ as function of $\sqrt{L}$, for several chemical potential profiles $V$.
The inset shows the occupations $\lambda_{n}(\bar{L})$, for $\bar{L}=800$. 
(c) Scaling collapse of $\lambda_0 L^\beta \times V L^\alpha$, for different system  sizes  $L$. 
Best fits to the data are compatible with $\alpha = 1$ and $\beta = -1/2$.
}
\label{fig: phase transition}
\end{figure}

{\it Model and Method ---}
We consider a tight-binding chain of HCB of size $L$ coupled to reservoirs at its edges, modeled by the Hamiltonian, ${\cal H}={\cal H}_{\text{C}}+\sum_{l}\left({\cal H}_{l}+{\cal H}_{\text{C},l}\right)$,
where ${\cal H}_{\text{C}} = - J \sum_{ \left\langle r,r' \right\rangle} \hat{b}_{r}^{\dagger} \hat{b}_{r'}$.
HCB at the $r$-th site are created (destroyed) by the operators $\hat{b}_{r}^{\dagger}$ ($\hat{b}_{r}^{ }$),
which fulfill the commutation relations $[\hat{b}_{r},\hat{b}_{r'}^{\dagger}]=\delta_{r,r'}(1-2\hat{b}_{r}^{\dagger}\hat{b}_{r})$.
Each of the two reservoirs ${\cal H}_{l}$  ($l=\text{L},\text{R}$) is a semi-infinite chain of HCB with hopping strength
$J_{l}$ and chemical potential $\mu_{l}$, held at zero temperature.
In the following, we will use $\mu=(\mu_{\text{L}}+\mu_\text{R})/2$ and $V=\mu_\text{L}-\mu_\text{R}$.
As shown in Fig.~\ref{fig: phase transition}(a),
the reservoirs are coupled (through ${\cal H}_{\text{C},l}$)
to the very left  ($r_{\rm L} \equiv 1$) and right site ($r_{\rm R} \equiv L$) of the chain, with coupling strength
$J_{\text{C},l}$.
In the following we make the simplifying assumption that the bandwidths of the reservoirs, $J_{l}$, are much larger than all other energy scales (`wide band limit'). In this limit, the coupling to each reservoir $l$ is completely determined by $\Gamma_l=\pi J_l^2 \rho_l$, the hybridization energy scale, where $\rho_l$ are the local densities of states of the reservoirs, taken to be energy independent.

The  Hamiltonian ${\cal H}$ possess a fermionic representation which can be obtained through the Jordan-Wigner mapping \citep{Lieb1961}, $\hat{b}_{r}^{\dagger}=e^{i\pi\sum_{r'=1}^{r-1}\hat{c}_{r'}^{\dagger}\hat{c}_{r'}}\hat{c}_{r}^{\dagger}$, where $\hat{c}_{r}^{\dagger}$ ($\hat{c}_{r}$) creates (annihilates) a spinless fermion at site $r$.  This yields a metallic chain in contact with  baths of spinless fermions held at chemical potentials  $\mu_{l=\text{L},\text{R}}^{}$.
As the Jordan-Wigner-transformed Hamiltonian is quadratic in its fermionic degrees of freedom, the nonequilibrium system admits an exact solution in terms of single-particle quantities. Thus, we employ the nonequilibrium Green function formalism to compute correlation functions and related observables. Steady-state observables can be obtained from the single-particle correlation-function matrix $\boldsymbol{\chi}\equiv\langle\hat{\boldsymbol{\Psi}}\cdot\hat{\boldsymbol{\Psi}}^{\dagger}\rangle $,
with $\hat{\boldsymbol{\Psi}}^{\dagger}=(\hat{c}_{1}^{\dagger},\ldots,\hat{c}_{N}^{\dagger})$,
which in turn is obtained from the Keldysh Green function as described
in Ref.~\cite{Puel-Chesi-Kirchner-Ribeiro-2021}. The method
allows us to obtain  mean values of quadratic observables
$\hat{O}=\hat{\boldsymbol{\Psi}}^{\dagger}\cdot\boldsymbol{O}\cdot\hat{\boldsymbol{\Psi}}$
from the relation $\langle \hat{O}\rangle =-\text{tr}\left[\boldsymbol{O}\cdot\boldsymbol{\chi}\right]$. 
Finally, the bosonic one-body density matrix $\rho_{r,r'}^{\text{B}}=\langle \hat{b}_{r}^{\dagger}\hat{b}_{r'}\rangle $ 
can be computed from the fermionic one, $\rho_{r,r'}^{\text{F}}=\langle\hat{c}_{r}^{\dagger}\hat{c}_{r'}\rangle = \delta_{r,r'}- \chi_{r',r} $, using the approach described, {\it e.g}., in Ref. \cite{PhysRevA.72.013604}.
One finds 
\begin{align}
\rho_{r,r'}^{B}	=\frac{1}{2}
\det\left[\sum_{i,j=1}^{r-r'}\left(2\rho_{j+r',i+r'-1}^{F}-\delta_{j+1,i}\right)\left|i\right\rangle \left\langle j\right|\right], \label{eq: rho_B}
\end{align}
for $r>r'$, 
and $\rho_{r',r}^{B} = \left ( \rho_{r,r'}^{B} \right )^*$. 
The eigenvectors of the matrix $\rho^B$ define the natural orbitals and the corresponding eigenvalues, $\lambda_n$ their occupations. Taking the $\lambda_n$ in decreasing order, the quasi-condensed state is characterized by 
a macroscopic occupation of its lowest orbital,
$\lambda_0\propto \sqrt{N_b} $, where $N_b \propto L$ in the macrocanonical ensemble \cite{doi:10.1063/1.1704196,PhysRevLett.42.3}.
On general grounds, the occupations behave as $\lambda_n (L\rightarrow \infty) \propto n^{-1/2}$ in the thermodynamic limit at zero temperature ($T$). In equilibrium, quasi-condensation is destroyed at non-zero temperature or the presence  of a localization potential. 


{\it Results --- } Figure \ref{fig: phase transition}(b) shows the occupation of the lowest natural orbital, $\lambda_0$, as a function of $L$, in the NESS  obtained for $V\neq 0$. 
In that case, $\lambda_0$ saturates with $L$ thus implying that the quasi-condensed state only exists for $V=0$. Nevertheless, the scaling $\lambda_0 \propto \sqrt{L}$ is still observed before the saturation scale is attained. 
The inset shows the scaling of 
$\lambda_n(L)$ with $n$, having fixed $L= 800$. 
For sufficiently large values of $n$, we find $\lambda_n\propto n^{-1/2}$, 
whereas, for small values, saturation ensues at finite values of $V$.
These findings establish that $V$ is a relevant perturbation, such as $T$ is in equilibrium. 
However, as  will be demonstrated, the NESS is fundamentally different from the finite-temperature state:  Its  critical behaviour, characterising the vicinity of  the unstable quasi-condensation fixed-point along the $V$ direction, turns out to be  different from the equilibrium case.   
Figure~\ref{fig: phase transition}(c) depicts the scaling collapse of $\lambda_0 L^{\beta} $ versus $VL^{\alpha}$ for different values of $L$. 
Best fits to the data are compatible with $\alpha=1$ and $\beta=-1/2$, which turn out to be the exact exponents, see below. 
For small $VL$ ($VL\le2$), this recovers the $V=0$ result, $\lambda_0 \propto \sqrt{L}$, whereas for $VL$ large,   $\lambda_0 \propto V^{-1/2} $. 

\begin{figure}
\hfill{}\includegraphics[width=0.90\columnwidth]{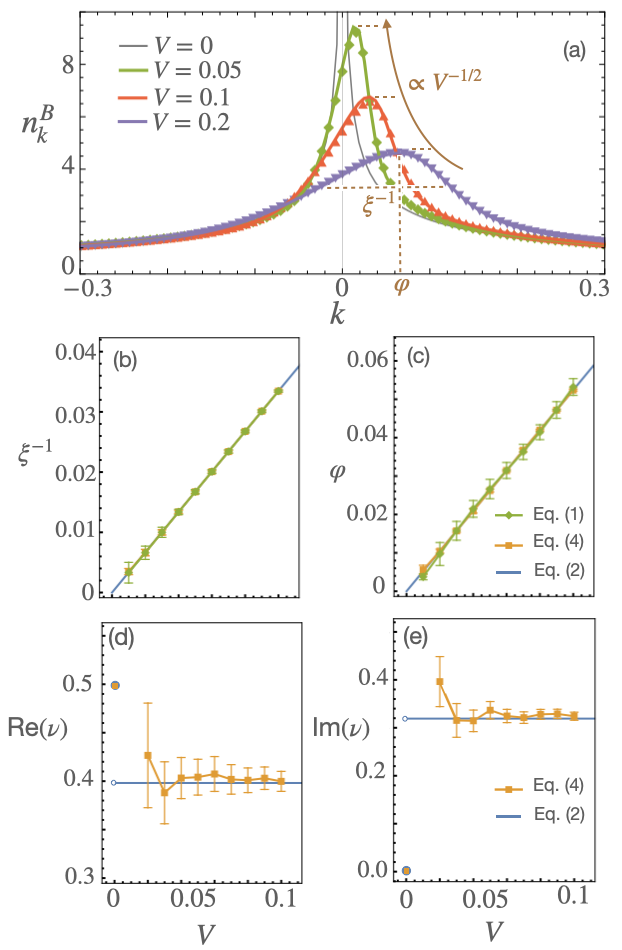}\hfill{} 
\caption{(a) Momentum distribution of the one-body density matrix for various values of the bias $V$. 
Comparison between the numerical and analytic results of the exponential decay length $\xi^{-1}$ (b), the momentum displacement $\varphi$ (c), and the real (d) and imaginary (e) parts of the power-law decay exponent $\nu$, defined in  Eq.~(\ref{eq:rho_B_2}).
Direct evaluation of Eq.~(\ref{eq: rho_B}) (green) is contrasted with results based on the asymptotic form in Eq.~(\ref{eq:rho_b_1}) (orange) and with the analytic expression of Eq.~(\ref{eq:rho_B_2}) (blue).  
} 
\label{fig: particle density}
\end{figure}

We now turn to the description of the NESS. 
For large system sizes, the state in the middle of the chain displays translational invariance and $\rho^B$ becomes diagonal in momentum space. In this case, the natural orbitals coincide with the momentum states. We label their occupations by $\lambda_n(L\gg 1) \rightarrow n^B_k$. 
Figure~\ref{fig: particle density}(a) depicts the Fourier transform of the bosonic occupation in momentum space in the middle of the chain. 
Here we also compare 
 numerical results (plot markers) for finite $L=800$ to analytical predictions (solid lines) valid for small $V$, discussed in detail below.  Clearly, both coincide for sufficiently small $V$.
The effect of $V$ is twofold: 
(i) the $1/\sqrt{k}$ divergence of $n^B_k$ at $V=0$ gets regularized at a scale  $1/\sqrt{V}$, and the curve acquires a characteristic  width to which we refer as $\xi^{-1}$,  illustrated in Fig.~\ref{fig: particle density}(a)  (for the blue curve); (ii) the maximum value of the peak shifts to finite momentum, denoted $\varphi$.   
Both quantities, together with a power-law exponent  $\nu$ [see  Eq.~(\ref{eq:rho_B_2}) below], characterize the departure from equilibrium of a quasi-condensate state, which is induced via a particle number bias.

A proper definition of  $\xi^{-1}$ and $\varphi$ is given in terms of the asymptotic dependence of the bosonic correlation function. In the limit $r-r'\to\infty$,
\begin{equation}
\rho_{r-r'}^{\text{B}}\simeq E \  \text{e}^{-\left|r-r'\right|/\xi-i\left(r-r'\right)\varphi}\left(r-r'\right)^{-\nu}, \label{eq:rho_B_2}
\end{equation}
where $E$ is a constant. As we will discuss in detail, this general dependence follows from taking the thermodynamic limit, which brings $\rho_{r,r'}^{\text{B}}$ to the form given in Eq.~(\ref{eq:rho_b_1}) and allows us to apply the Fisher-Hartwig conjecture for Toeplitz matrices. Interestingly, the asymptotic behavior of $\rho_{r-r'}^{\text{B}}$ displays power-law corrections on top of the exponential decay, where $\nu$ is a  complex-valued critical exponent. The behavior of $\xi^{-1}$ and $\varphi$ with $V$, obtained by fitting the numerical $\rho_{r,r'}^{\text{B}}$ to Eq.~(\ref{eq:rho_B_2}),  is given by the green dots in Figs.~\ref{fig: particle density}(b) and (c), respectively. These results agree (within the error bars) with our analytical formulae, given below. Furthermore, in Figs.~\ref{fig: particle density}(d) and (e) we examine the power-law exponent $\nu$. While a finite value  $\nu=1/2$ is expected in equilibrium when $\xi^{-1}=\varphi=0$, 
our analytic results show that $\nu$ is discontinuous at $V=0$ and assumes a $V$-independent constant for any non-vanishing value of the bias, {\itshape e.g., $V\neq 0$}. This discontinuity only occurs in the thermodynamic limit. Our numerical results show that  $\nu$ indeed remains  constant for a finite chain at a sufficiently large $V$ but will acquire strong finite size corrections as $V$ is reduced, see Figs.~\ref{fig: particle density}(d) and (e). Power-law corrections on top of the exponential decay are hard to determine based on Eq.~(\ref{eq: rho_B}). Results shown in both panels (d) and (e) are therefore obtained  by a numerical evaluation of Eq.~(\ref{eq:rho_b_1}) below, that allows accessing much larger system sizes. Nevertheless, we confirmed (see SM \cite{suppmaterial}) that the numerical results obtained with Eq.~(\ref{eq: rho_B}) are fully compatible with the analytic asymptotic form.   

These findings constitute the main non-technical results of our work. In what follows, we explain the method used to obtain our numerical results and derive Eq.~(\ref{eq:rho_B_2}), including an explicit expression for $\xi,\varphi$ and $\nu$.

{\it Single-particle correlations --- }  The numerical evaluation of single-particle correlators is most conveniently performed in the fermionic representation which leads to the non-Hermitian single-particle operator $\boldsymbol{K}=\boldsymbol{H}_{\mathrm{C}}-i \sum_{l=\mathrm{L}, \mathrm{R}} \boldsymbol{\gamma}_{l} $, with the Hamiltonian of the chain $\boldsymbol{H}_{\mathrm{C}}=- J \sum_{r=1}^{L-1} \ket{r}\bra{r+1} + \text{h.c.}$, and where $|r\rangle$ is a single-particle state. The hybridization matrices of each reservoir are $\boldsymbol{\gamma}_{l}=\Gamma_{l}\left|r_{l}\right\rangle\left\langle r_{l}\right|$. We assume that $\boldsymbol{K}$ is diagonalizable, having right and left eigenvectors $|\alpha\rangle$ and $\langle\tilde{\alpha}|$, with associated eigenvalues $\lambda_{\alpha}$.
The single-particle correlation-function matrix  $\boldsymbol{\chi}$ is given by \cite{PhysRevA.72.013604,Puel-Chesi-Kirchner-Ribeiro-2019, Puel-Chesi-Kirchner-Ribeiro-2021}
\begin{eqnarray}
\boldsymbol{\chi}&=&\frac{1}{2}+\sum_{l=\mathrm{L}, \mathrm{R}} \sum_{\alpha \beta}|\alpha\rangle\langle\beta| \times \nonumber \\
& &\left\langle\tilde{\alpha}\left|\left[\gamma_{l} I_{l}\left(\lambda_{\alpha}, \lambda_{\beta}^{*}\right)-\hat{\gamma}_{l} I_{l}\left(-\lambda_{\alpha},-\lambda_{\beta}^{*}\right)\right]\right| \tilde{\beta}\right\rangle,
\label{eq:chi}
\end{eqnarray}
where $I_{l}\left(z, z^{\prime}\right)=-\frac{1}{\pi} \frac{g\left(z- \mu_{l}\right)-g\left(z^{\prime}- \mu_{l}\right)}{z-z^{\prime}}$ with $g(z)=$ $\ln (-i \operatorname{sgn}[\operatorname{Im}(z)] z)$. 
The matrix $\boldsymbol{\chi}$ is then used to calculate the bosonic one-body density matrix as in Eq.~(\ref{eq: rho_B}).

\begin{figure}
\hfill{}\includegraphics[width=0.90\columnwidth]{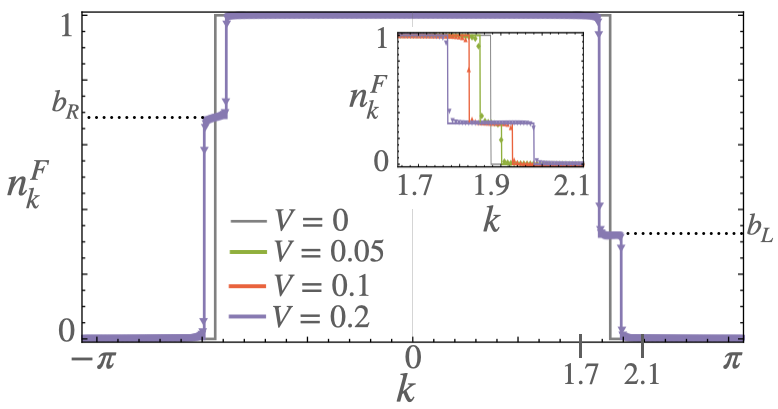}\hfill{} 
\caption{  
Momentum distribution function of the Jordan-Wigner fermions for $V=0$ and $V=0.2$.
Insert shows the double-step structure for different values of $V$. The step width is $V/v_\text{F}$ and the heights, $b_L$ and $b_R$, depend on the chain-lead couplings. 
The  symbols are obtained by Fourier transformation of Eq.~(\ref{eq:chi}) and the lines are analytic predictions of Ref.~\cite{Ribeiro.17}. 
}
\label{fig: n_f and nu}
\end{figure}

Alternatively, one can obtain the fermionic one-body density matrix analytically.
Noticing that  in the bulk of an infinite chain the Fourier transform of $\rho^F_{r,r'} = \rho^F_{r-r'}$ becomes the momentum occupation number $n^F_k$,
an explicit expression of $n^F_k$ can be obtained when the energy dispersion is represented by a linear $k$ dependence near the Fermi points, provided $V\ll J$  \cite{Ribeiro.17}. 
In this case, $n^F_k$ assumes the double-step structure illustrated in Fig.~\ref{fig: n_f and nu} (see \cite{Ribeiro.17, suppmaterial}). Each double-step has a width of $V/v_{\text F}$, with the Fermi velocity $v_{\text F}=2J\sin\left(k_{\text F}\right)$, and is centered around the average Fermi momentum $k_F=\arccos[-\left(\mu_{L}+\mu_{R}\right)/4J]$. The occupations of the left (right), $b_R$ ($b_L$) steps depend on the couplings to the reservoirs and on the Fermi velocity. For details, see Refs.\ \cite{Ribeiro.17, suppmaterial}.
%
%
Expressed in terms of $n^F$, the asymptotic limit of $\rho^B$ assumes the form of a Toeplitz matrix. Explicitly, one finds \cite{suppmaterial}   
\begin{align}
\rho_{r-r'}^{\text{B}}  =\frac{1}{2}\det\left[\sum_{i,j=1}^{r-r'}\int\frac{dk}{2\pi}\left(2n_{k}^{F}-1\right)\text{e}^{ik\left(i-j-1\right)}\left|i\right\rangle \left\langle j\right|\right].
 \label{eq:rho_b_1}
\end{align}

Equation (\ref{eq:rho_B_2}) was obtained using the Fisher-Hartwig conjecture for Toeplitz matrices,  giving the asymptotic behavior of Eq.~(\ref{eq:rho_b_1}) in the limit $r-r'\to\infty$ (see \cite{suppmaterial}).
The correlation length $\xi^{-1}$, the displacement momentum $\varphi$, and the real and imaginary parts of the power-law exponent $\nu$ are explicitly given by
\begin{align}
\xi^{-1} & =-\frac{1}{2\pi} \frac{\left|V\right|}{v_{\text F}} \log\left(\left|1-2b_{L}\right|\left|1-2b_{R}\right|\right), \nonumber\\
\varphi & =\frac{1}{4} \frac{ V }{v_{\text F}} \left[\text{sign\ensuremath{\left(1-2b_{L}\right)}}-\text{sign\ensuremath{\left(1-2b_{R}\right)}}\right],
\nonumber\\
\text{Re}\left(\nu\right) & =\frac{1}{2}-\frac{1}{2\pi^{2}}\left[\log^{2}\left(\left|1-2b_{L}\right|\right)+\log^{2}\left(\left|1-2b_{R}\right|\right)\right],\nonumber\\
\text{Im}\left(\nu\right) &  =\frac{\text{sign}\left(V\right)}{2\pi}\left[\frac{\log\left(\left|1-2b_{R}\right|\right)}{\text{sign\ensuremath{\left(1-2b_{R}\right)}}}-\frac{\log\left(\left|1-2b_{L}\right|\right)}{\text{sign\ensuremath{\left(1-2b_{L}\right)}}}\right].\nonumber 
\end{align}
The constant $E$ in Eq.~(\ref{eq:rho_B_2}) can also be obtained explicitly as a function of $V/v_\text{F}$, $b_\t{L}$ and $b_\t{R}$, and is given in the SM \cite{suppmaterial}. 

{\it Discussion --- }  The numerical and analytical results presented so far  allow us  to address the fate of quasi-condensation at finite bias and contrast it with  what happens when turning on the heat at zero bias. 
Both, non-zero $V$ and non-zero $T$ are relevant perturbations which destroy quasi-condensation 
and lead to an exponential decay of correlations with distance. 
This is reflected in similar finite-size scaling behavior of $\lambda_{0}$ vs. $V$, see
Fig.~\ref{fig: phase transition}(c), and $T$, obtained in \cite{Rigol2005}.
There are, however, clear differences in how this destruction occurs in both cases. While the equilibrium correlations at finite $T$ decay exponentially, the out of equilibrium decay, as we have shown, is characterized by additional power law corrections on top of the exponential decay, see Eq.~(\ref{eq:rho_B_2}).
These differences will be most apparent at the short to intermediate range where these power-law corrections are sizable and are, e.g., also reflected in the behavior of the mutual information of the Jordan-Wigner fermions~\cite{Ribeiro.17, Vidal.03}. 
%
Yet another difference between the two cases concerns  the steady-state realized in the  $V\to0$ limit. 
In the thermodynamic limit, $\abs{V}\to0^+$ results in a divergence of $\xi$, thus recovering the quasi-condensed state. However, the power-law scaling characterising this state  depends on the coupling to the leads and is in general different from the  $\nu=1/2$ observed in equilibrium. 

An important direction concerning future work is the stability of our findings with respect to interactions. It will be interesting to address the effect of a relaxation of the hard-core constraint and the resulting softening of occupation numbers.  Another open question 
regards the role of dimensionality. True condensation occurs in two and higher dimensional systems in equilibrium but its fate out of equilibrium has so far remained unclear. In particular, it would be worthwhile to understand if this out-of-equilibrium steady state is fundamentally different from its thermal counterpart, in analogy to what happens in 1D.

	\begin{acknowledgments}
		\vspace{10pt}
		\textit{Acknowledgments --} 
		We thank Beijing Computational Science Research Center (CSRC) for providing access to the Tianhe-2JK cluster where the calculations were performed.
		S.\,C.\, acknowledges support from the National Science Association Funds (Grant No. U1930402) and NSFC (Grants No. 11974040 and No. 12150610464). S.\,K.\, acknowledges support by the Ministry of Science and Technology, Taiwan (grant No. MOST 111-2634-F-A49-007) and the Featured Area Research Center Program within the framework of the Higher Education Sprout Project by the Ministry of Education (MOE) in Taiwan. P.\,R.\,  acknowledges support by FCT through Grant No. UID/CTM/04540/2019.
	\end{acknowledgments}

\bibliographystyle{apsrev4-1}
\bibliography{refs}

\clearpage
\pagebreak 
\newpage

\begin{widetext}
\begin{center}
\textbf{\large{}\textemdash{} Supplemental Material \textemdash{}}
\par\end{center}{\large \par}
\begin{center}
\textbf{\large{}Fate of the Quasi-condensed State for bias-driven Hard-core Bosons in one Dimension}
\par\end{center}{\large \par}
\begin{center}
\textbf{T. O. Puel$^{1,2}$, S. Chesi$^{3,4}$, S. Kirchner$^{5,6}$, P. Ribeiro$^{7,3}$}\\
$^{1}$Zhejiang Institute of Modern Physics, Zhejiang University,
Hangzhou, Zhejiang 310027, China~\\
$^{2}$Department of Physics and Astronomy, University of Iowa, Iowa City, Iowa 52242, USA~\\
$^{3}$Beijing Computational Science Research Center, Beijing 100193,
China~\\
$^{4}$Department of Physics, Beijing Normal University, Beijing 100875,
China~\\
$^{5}$Department of Electrophysics, National Yang Ming Chiao Tung University, Hsinchu 30010, Taiwan\\
$^{6}$Center for Emergent Functional Matter Science, National Yang Ming Chiao Tung University, Hsinchu 30010, Taiwan\\
$^{7}$CeFEMA, Instituto Superior T\'{e}cnico, Universidade de Lisboa
Av. Rovisco Pais, 1049-001 Lisboa, Portugal
\end{center}

\begin{description}
\item [{Summary}] Below we provide technical details and  numerical results supplementing the conclusions from the main text.
\end{description}
\end{widetext}

\beginsupplement

\section{Fermionic-particle density\label{app: fermionic-particle density}}

The fermionic-particle density, computed in Ref. \citep{Ribeiro.17}, is
\begin{equation}
n_{k}^{F}=\begin{cases}
1 & 0<k<\theta_{1}\\
b_{L} & \theta_{1}<k<\theta_{2}\\
0 & \theta_{2}<k<\theta_{3}\\
b_{R} & \theta_{3}<k<\theta_{4}\\
1 & \theta_{4}<k<2\pi
\end{cases},\label{eq: nFk regions}
\end{equation}
with the momenta where $n^\text{F}$ is discontinuous given by
\begin{align}
\theta_{1} & =k_{F}-\frac{\Delta}{2},\qquad\theta_{2}=k_{F}+\frac{\Delta}{2},\nonumber \\
\theta_{3} & =2\pi-k_{F}-\frac{\Delta}{2},\qquad\theta_{4}=\pi-k_{F}+\frac{\Delta}{2},\label{eq: nkF discontinuities}
\end{align}
where $\Delta=V/v_\text{F}$. 
The values of $b_{R}$ and $b_{L}$ are obtained
from
\begin{equation}
b_{R}=-\frac{\left(1-\gamma_{L}\right)}{\gamma_{L}\gamma_{R}-1},\quad b_{L}=\gamma_{R}b_{R},
\end{equation}
in which
\begin{equation}
\gamma_{l}=\frac{\left(\Gamma_{l}/J\right)^{2}-\left(2/J\right)\sin\left(k_{F}\right)\Gamma_{l}+1}{\left(\Gamma_{l}/J\right)^{2}+\left(2/J\right)\sin\left(k_{F}\right)\Gamma_{l}+1},\quad l=L,R.
\end{equation}

\section{Toeplitz determinant and Fisher-Hartwig conjecture\label{app: Toeplitz determinant}}

In the thermodynamic limit the matrix $\rho^{F}_{i,j}$, for $i$ and $j$ in a segment in the middle of the chain, becomes translationally invariant and assumes the Toeplitz form $\rho^{F}_{i,j} = \rho^{F}_{i-j} $.  Its  asymptotic behavior is given by Eq.~(\ref{eq:rho_b_1}). 
Physical quantities related to $\rho^{F}$ thus require the  evaluation  of the determinant of Toeplitz matrices, which we perform in the following using the Fisher-Hartwig conjecture. 

\subsection{Toeplitz matrix}

Consider a Toeplitz matrix $T_{n}\left[\phi\right]=\sum_{q,l=1}^{n}\phi_{q-l}\left|q\right\rangle \left\langle l\right|$, generated by a function of the form
\begin{equation}
\phi_{q-l}=\int_{0}^{2\pi}\frac{d\theta}{2\pi}\phi\left(\theta\right)\text{e}^{-i\left(q-l\right)\theta},\label{eq: general Toeplitz generating function}
\end{equation}
where the non-analyticities of $\phi(\theta)$ are assumed to consist only of discontinuities. In this case, it can be decomposed in the form \cite{Basor.94} 
\begin{equation}
\phi\left(\theta\right)=b\left(\theta\right)\prod_{r=1}^{R}\text{e}^{-i\beta_{r}\left(\pi-\left(\theta-\theta_{r}\right)\right)},\label{eq: discontinuities in the generating function}
\end{equation}
where $\theta_{r}$ are the discontinuity points. 
Comparing  Eq.~(\ref{eq:rho_b_1}) in the main text
and Eq. (\ref{eq: general Toeplitz generating function}) we identify
\begin{equation}
\phi\left(\theta\right)\rightarrow\phi\left(k\right)=\left[1-2n_{k}^{F}\right]e^{-ik},
\end{equation}
with $n_{k}^{F}$ given in Eq. (\ref{eq: nFk regions}). The discontinuities $\theta_{r}$ of $n_{k}^{F}$ are defined in Eq.~(\ref{eq: nkF discontinuities}), for $r=1,2,3,4$. 
In order to identify the coefficients $\beta_{r}$ we impose
\begin{equation}
\text{e}^{ \ln\left[1-2n_{k}^{F}\right]-ik }=\text{e}^{ V_{0}-i\sum_{r}\beta_{r}\left(\pi-\left(k-\theta_{r}\right)\right) },
\end{equation}
where we defined $b\left(k\right)\equiv\exp\left[V_{0}\right]$ in Eq.~(\ref{eq: discontinuities in the generating function}), to be valid for each of the  continuous regions of Eq.~(\ref{eq: nFk regions}).
In between the regions, the exponents of both sides have to coincide up to a constant $2 \pi i n_j$ with $n_j\in\mathbb{Z}$ defined in region $j$. Note that there are only four regions since $0<k<\theta_{1}$
and $\theta_{4}<k<2\pi$ are connected by periodicity. 

Equating the coefficients multiplying $k$ on both sides we obtain
\begin{equation}
\sum_{r=1}^{4}\beta_{r}=-1,
\end{equation}
which leaves us with only three independent variables, $n_{1}$, $n_{2}$, and $n_{3}$, that must satisfy
\begin{align}
\beta_{1}= & n_{2}-n_{1}+\frac{i\log\left(1-2b_{R}\right)}{2\pi}+\frac{1}{2},\\
\beta_{2}= & n_{3}-n_{2}-\frac{i\log\left(1-2b_{R}\right)}{2\pi},\\
\beta_{3}= & n_{4}-n_{3}+\frac{i\log\left(1-2b_{L}\right)}{2\pi},\\
\beta_{4}= & -1-\beta_{1}-\beta_{2}-\beta_{3},
\end{align}
and
\begin{align}
V_{0}= & \frac{\Delta}{2\pi}\left[\log\left(1-2b_{\text{L}}\right)+\log\left(1-2b_{\text{R}}\right)\right]\nonumber \\
 & -i\left[\pi+2\left(n_{1}-n_{3}-1\right)k_{\text{F}}- \right. \nonumber\\
 & \left.
 \Delta\left(n_{1}-n_{2}+n_{3}-n_{4}+1\right)\right].
\end{align}
The values of the integers $n_j$ are determined in the following, using the formulation of the Fisher-Hartwig conjecture in Ref.~\cite{Basor.94}.  

\subsection{Fisher-Hartwig conjecture}

The Fisher-Hartwig conjecture states that
\begin{equation}
\det T_{\ell}\left[\phi\right]\simeq E\text{e}^{\ell V_{0}}\ell^{-\sum_{r}\beta_{r}{}^{2}},\quad \ell\rightarrow\infty,\label{eq: Fisher-Hartwig conjecture}
\end{equation}
where $E$ is an $\ell$-independent constant evaluated below. In the following, we chose the values of $n_j$ that maximize Eq.~(\ref{eq: Fisher-Hartwig conjecture}). 
The dependence of $V_{0}$ on $n_{1},n_{2},n_{3}$ does not affect the absolute value of $\det T_{\ell}$. 
Therefore, the determination of $n_j$ is obtained by minimizing  $\text{Re}\left[\sum_{r}\beta_{r}{}^{2}\right]$.
The result depends on $\text{sign}\left(\Delta\right)$ 
and whether $\left(1-2b_{L/R}\right)$ is positive or negative. After this procedure, the expressions for $\beta_j$ write
\begin{align}
\beta_{1}= & -\frac{1}{2}-i \  \text{sign}\left(\Delta\right)\frac{\text{sign}\left(1-2b_{R}\right)\log\left(\left|1-2b_{R}\right|\right)}{2\pi},\\
\beta_{2}= & -\frac{1}{2}-\beta_{1},\\
\beta_{3}= & -i \ \text{sign}\left(\Delta\right)\frac{\text{sign}\left(1-2b_{L}\right)\log\left(\left|1-2b_{L}\right|\right)}{2\pi},\\
\beta_{4}= & -1-\beta_{1}-\beta_{2}-\beta_{3},
\end{align}
and
\begin{align}
V_{0}= & \frac{\left|\Delta\right|}{2\pi}\log\left(\left|1-2b_{L}\right|\left|1-2b_{R}\right|\right),\nonumber \\
 & +i\left[\pi+\frac{\Delta}{4}\left[\text{sign\ensuremath{\left(1-2b_{L}\right)}}-\text{sign\ensuremath{\left(1-2b_{R}\right)}}\right]\right].
\end{align}
Using Ref.~\cite{Basor.94}, it follows that the amplitude $E$ in Eq. (\ref{eq: Fisher-Hartwig conjecture}) is obtained from the expression
\begin{align}
E=\prod_{1\leq r\neq s\leq4}\left(1-\text{e}^{i\left(\theta_{s}-\theta_{r}\right)}\right)^{\beta_{r}\beta_{s}} \prod_{k=4}^{4}G\left(1+\beta_{k}\right)G\left(1-\beta_{k}\right),
\end{align}
where $G$ is the Barnes G-function,
\begin{equation}
G\left(1+z\right)=\left(2\pi\right)^{z/2}\text{e}^{-\left(z+\left(\gamma+1\right)z^{2}\right)/2}\prod_{k=1}^{\infty}\left(\frac{1+z}{k}\right)^{k}\text{e}^{-z+z^{2}/\left(2k\right)},
\end{equation}
and $\gamma$ is the Euler constant.

\subsection{Bosonic single-particle matrix}

Recasting the results of the previous sections, 
we evaluate Eq.~(\ref{eq:rho_b_1}) in the main text and obtain
\begin{equation}
\rho_{\left|r-r'\right|}^{\text{B}}=\frac{E}{2}\text{e}^{\left|r-r'\right|V_{0}} \left|r-r'\right|^{-\nu}.
\end{equation}
The correlation function and the momentum shift, defined in the main text, can be identified as  $V_{0}=-\xi^{-1}+i\varphi$ 
and $\nu=\sum_{r}\beta_{r}{}^{2}$. The corresponding expressions are given explicitly in the main text. 

\begin{figure}[b]
\hfill{}\includegraphics[width=0.80\columnwidth]{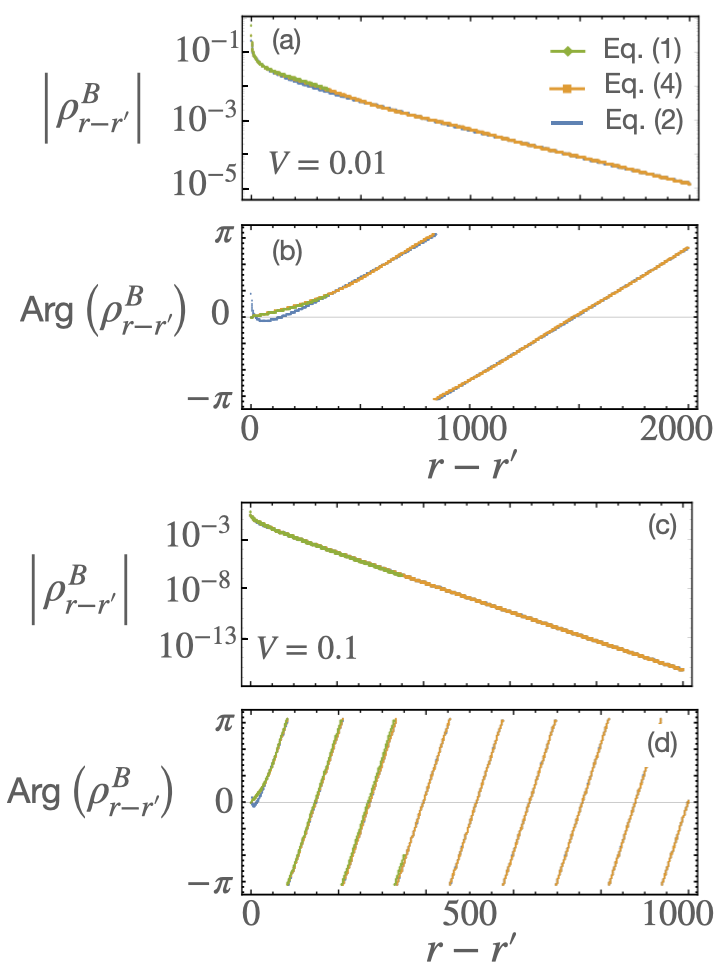}\hfill{}
\caption{
(a) and (b) are the particle density for $V=0.01$, while (b) and (c) are the particle density for $V=0.1$. 
}
\label{fig:S1}
\end{figure}

Fig.~\ref{fig:S1} shows a real-space comparison of the asymptotic result, obtained here using the same color coding as in the main text. 
Although the results based on the different methods are compatible with each other, for small values of $V$, it is difficult to access the asymptotic regime for large $r-r'$  using Eq.~(\ref{eq: rho_B}). This difficulty explains the growing error bars and why we were not able to provide full numerical results of small $V$ in Fig.~\ref{fig: particle density}.

\end{document}